\documentstyle[12pt]{article}
\def\be{\nopagebreak[3]\begin{equation}}
\newcommand{\ee}{\end{equation}}
\def\ba{\begin{array}}
\def\ea{\end{array}}
\def\nl{\vspace{1pc} \\}

\newcommand{\dsp}{\displaystyle}
\newcommand{\scs}[1]{{\scriptscriptstyle #1}}
\newcommand{\im}{{\rm Im}\,}

\newcommand{\tr}{{\rm tr}\,}
\newcommand{\eq}[1]{Eq.~(\ref{#1})}
\newcommand{\pvint}{\int\!\!\!\!\!\! - }
\renewcommand{\d}{\partial}
\renewcommand{\ss}[1]{{\hbox{}_{#1}}}

\newcommand{\N}{\scs N}
\newcommand{\R}{\scs R}
\renewcommand{\S}{\scs S}

\newcommand{\sn}{{\rm sn\,}}
\newcommand{\dn}{{\rm dn\,}}
\newcommand{\cn}{{\rm cn\,}}
\newcommand{\qt}{\tilde{q}}

\begin{document}
\begin{titlepage}
\begin{flushright}
NBI-HE-93-57\\
October 5 ,1993
\end{flushright}
\vspace*{36pt}
\begin{center}
{\huge \bf
Wilson loop on a sphere
}\end{center}
\vspace{2pc}
\begin{center}
 {\Large D.V. Boulatov}\\
\vspace{1pc}
{\em The Niels Bohr Institute\\
University of Copenhagen\\
Blegdamsvej 17, 2100 Copenhagen \O \\
Denmark}
\vspace{2pc}
\end{center}
\begin{center}
{\large\bf Abstract}
\end{center}
We give the formula for a simple Wilson loop on a sphere
which is valid for an arbitrary QCD$_2$ saddle-point $\rho(x)$:
\mbox{$W(A_1,A_2)=\oint \frac{dx}{2\pi i}
\exp(\int dy \frac{\rho(y)}{y-x}+A_2x)$}. The strong-coupling-phase
solution is investigated.
\vfill
\end{titlepage}

\section{Introduction}

$QCD_2$ has recently drawn much
attention because of the new insights brought about by the Gross-Taylor
reinterpretation of its partition function as the sum over branched
coverings  \cite{Gross}. The similar interpretation exists for Wilson
loop operators on the plane as well \cite{Kos}.
The topological nature of $QCD_2$ was noticed
long ago by A.A.Migdal \cite{Mig} who proposed to use a character expansion of
Boltzman weights to manifest it. One can always integrate over a link
variable by using only the orthogonality of group elements.
Following this guideline, the partition
function of $QCD_2$ on an arbitrary 2-manifold was obtained in ref.~\cite{Rus1}
in the form of
a sum over irreps of a gauge group. Recently, on
the sphere, the third order phase transition with respect to the area
was found \cite{DougKaz} (see also \cite{Rus2}, where the
weak coupling partition
function was calculated). Wilson loop operators on the plane were
investigated in refs.~\cite{KazKos} and shown to display stringy
features, in particular, to obey the area low. However, in the weak
coupling phase on the sphere their behavior is completely different
\cite{DougKaz}. In the present paper, we give the general solution for a
simple loop
establishing a bridge between results of refs.~\cite{KazKos} and
\cite{DougKaz}. Having known the answer, non-trivial Wilson loops can be
found by solving the renormalized
Makeenko-Migdal equation adopted for the sphere
\cite{Rus3}.

\bigskip

For reader's convenience, let us repeat known results from
Refs.~\cite{Rus1,DougKaz,Rus2} which we shall need later.
It will also allow us to introduce a notation.

If the Boltzmann weight, as a function of a holonomy along a plaquette, is
taken in the form of the $U(N)$ heat kernel, it is reproduced
after the integration over internal gauge variables, and one can easily
obtain the partition function for a disc (as a function of an area, $A$,
and a holonomy along the boundary,
$U$) in the form of the sum over $U(N)$ representations

\be
Z_1(g^2A,U)=\sum_\R d_\R\chi_\ss{R}(U)\exp(-\frac{g^\scs{2}A}{2N}C_R)
\label{Z1=}
\ee
where $\chi_\ss{R}(U)$ is the character of an $R$'th irrep;
$d_\R=\chi_\ss{R}(I)$ is its dimension:

\be
d_\R = \prod_{1\leq i < j \leq N} \Big(1+\frac{n_i-n_j}{j-i}\Big)
\label{dim}
\ee
$C_\R$ is the value of the second Casimir
operator:

\be
C_\R = \sum_{i=1}^{N} n_i(n_i +N+1-2i)
\label{Casimir}
\ee
where $n_i$ are highest weight components of $R$,
$R\equiv[n_1,\ldots,n_\N]$, {\em i.e.} the numbers of boxes in rows of
the Young table ($n_1\geq n_2 \geq \ldots \geq n_\N$). We are
considering $U(N)$, therefore $n_\N$ is unrestricted.

The gauge
coupling constant $g^\scs{2}$ can be absorbed into the area, $A$,
(in what follows, we put $g^\scs{2}=1$).
For $A=0$, \eq{Z1=} becomes the group
$\delta$-function: $Z_1(0,U)=\delta(U,1)$.

The partition function for a closed surface of a genus $p$
can be represented as the sum \cite{Rus1}:

\be
Z_p(A)=\sum_{R} d^{2(1-p)}\exp(-\frac{A}{2N}C_R)
\label{Zp=}
\ee

In the spherical case, $p=0$, the sum becomes divergent at small areas.
Hence, in the large $N$ limit, a non-trivial saddle-point should exist.
In this case we can introduce the continuous function \cite{Rus2}

\be
h(x)=\lim_{N\to\infty} \frac{1}{N}\Big(i - \frac{N}{2} - n_i\Big);
\hspace{1cm}
x=\frac{i}{N}-\frac{1}{2}
\ee
then the saddle-point equation takes the well known form

\be
\frac{A}{2}h=\pvint dy \frac{\rho(y)}{h-y}
\label{spe1}
\ee
Here,

\be
\rho(h)=\frac{\d x}{\d h}
\label{rho}
\ee
obeys the inequality

\be
\rho(h)\leq 1
\label{rho<1}
\ee

If \eq{rho<1} is ignored, the solution to \eq{spe1} is the well known
semi-circle distribution

\be
\rho(h)=\frac{1}{\pi}\sqrt{A-\frac{A^2h^2}{4}}
\label{semicirle}
\ee
which is obviously valid for small areas: $A<\pi^2$. For larger areas
($A>\pi^2$), the inequality (\ref{rho<1}) is crucial and Douglas and
Kazakov \cite{DougKaz} have found in this case for the function

\be
f(z) = \int dy \frac{\rho(y)}{z-y}
\label{f}
\ee
the following answer

\be\ba{rcl}
f(z) &=&
\frac{A}{2}z +\frac{1}{z}{\dsp \int_{-b}^{+b}} \frac{dy}{1-\frac{y^2}{z^2}}
\sqrt{\frac{(a^2-z^2)(b^2-z^2)}{(a^2-y^2)(b^2-y^2)}}=
\vspace{1pc}
\\&&
\frac{A}{2}z+\frac{2}{az}\sqrt{(a^2-z^2)(b^2-z^2)}
\Pi_1\Big(-\frac{b^2}{z^2},\frac{b}{a}\Big)
\ea
\label{DKsol}
\ee
where $\Pi_1(x,k)$ is the complete elliptic integral of the third kind
with the modulus $k=\frac{b}{a}$.
The density is the imaginary part of this function on the cut,
\mbox{$\rho(z)=\frac{1}{\pi} \im f(z)$}, and has the form

\be
\rho(z)=\left\{\ba{ll}
-\frac{2}{\pi az}\sqrt{(a^2-z^2)(z^2-b^2)}
\Pi_1\Big(-\frac{b^2}{z^2},\frac{b}{a}\Big) & \mbox{ for $-a<z<-b$}\\
1 & \mbox{ for $-b<z<b$}\\
\frac{2}{\pi az}\sqrt{(a^2-z^2)(z^2-b^2)}
\Pi_1\Big(-\frac{b^2}{z^2},\frac{b}{a}\Big) & \mbox{ for $b<z<a$}
\ea\right.
\ee
The parameters are to be determined from the equations\footnote{Working
with elliptic functions we always use conventions adopted in the book
\cite{Erd}.}

\be
a(2E-k'^2K)=1 \hspace{1.5cm} aA=4K
\label{eqspar}
\ee
At the critical value $A_c=\pi^2$, a third order phase transition takes
place.

\par
Another interesting quantity is the Wilson loop average on a sphere:

\be\ba{rcl}
W(A_1,A_2) &=& \frac{1}{Z_0(A_1+A_2)} {\dsp \int} dU Z_1(A_1,U) Z_1(A_2,U^+)
\frac{1}{N}\tr U=
\vspace{1pc}
\\
&&\frac{1}{NZ_0(A_1+A_2)} {\dsp \sum_{\R,\S}} d_\R d_{\scs S}
\langle R, f|S \rangle
\exp(-\frac{A_1}{2N}C_R-\frac{A_2}{2N}C_S)
\ea
\label{W}
\ee
where $\langle R, f|S \rangle = 0$ or $1$ is the multiplicity of an irrep
$S$ in the tensor product of $R$ with the fundamental representation
$f$. Douglas and Kazakov \cite{DougKaz} gave for $W$ the expression

\be
W(A_1,A_2) = \sqrt{\frac{A_1+A_2}{A_1A_2}}
J_1\Big(\sqrt{\frac{4A_1A_2}{A_1+A_2}}\Big)
\label{W=J1}
\ee
valid in the weak coupling (small areas) phase and wrote after that ``It
would be very interesting to calculate the Wilson loop in the strong
coupling phase'', which is our goal in the present paper.

\section{Wilson loop for an arbitrary Young table}

If the sum over irreps is dominated by a particular representation $R$,
\eq{W} can be written as

\be
W(A_1,A_2)= \frac{1}{N}\sum_{\S} \frac{d_\S}{d_\R} \langle R, f|S \rangle
\exp(-\frac{A_2}{2N}(C_S-C_R))
\label{Wsp}
\ee
Obviously, $W(A_1,0)=1$, because of the identity

\be
\frac{1}{N} \sum_{\S} \frac{d_\S}{d_\R} \langle R, f|S \rangle = 1
\ee
The equality $W(0,A_2)=1$ is less trivial in this non-symmetric
representation and should hold dynamically ({\em i.e.} depending on an
actual form of the saddle-point $R$). If the number of boxes in the
Young table of $R$ is of the order $N^2$, the formula
(\ref{Wsp}) is really symmetric. However, if the saddle-point is
dominated by the trivial representation, but $S=0$ and $R=\bar{f}$,
it is obviously incorrect, because $Z_0=1$.
Nevertheless, our final answer will be free from this loop-hole.

The sum in \eq{Wsp} goes over all Young tables $S$ which can be obtained
from the one for the saddle-point $R$ by adding one box.
It is possible in rows $i$ where $n_i<n_{i-1}$.
Therefore, we can introduce the density of vacancies

\be
\eta(h) = \left\{\ba{ll}
1 & \mbox{ if $\rho(h)<\frac{1}{2}$}\\
1-\rho(h) & \mbox{ if $\frac{1}{2}<\rho(h)<1$}
\ea \right.
\label{eta}
\ee
and, like it was done in \eq{spe1} and (\ref{rho}),
rewrite \eq{Wsp} as the integral

\be
W(A_1,A_2) = \int dh \eta(h) D(h) e^{A_2h}
\ee
where we have used that, for Young tables of $S$ and $R$ differing by one
box in the $k$'th row,

\be
\frac{1}{N}(C_S-C_R)=\frac{1}{N}(2n_k + N+1-2k)
\to -2h(x)
\ee
$D(h)$ is the large $N$ limit of $d_\S/d_\R$:

\be\ba{l}
D\Big(\frac{k}{N}-\frac{1}{2}\Big)=
\frac{\prod_{i\neq k}(n_k+1-k-n_i+i)}{\prod_{i\neq k}(n_k-k-n_i+i)}=
\exp {\dsp \sum_{i\neq k} }
\log\Big(1-\frac{1/N}{h(\frac{k}{N}-\frac{1}{2})-
h(\frac{i}{N}-\frac{1}{2})}\Big)
\ea
\label{D}
\ee

We should be accurate in calculating the sum in the exponent. Let
\mbox{$M=O(\sqrt{N})$}. If $\rho(h)<\frac{1}{2}$,
we can substitute, for $|i-k|<M$,

\be\textstyle
N[h(\frac{i}{N}-\frac{1}{2})-h(\frac{k}{N}-\frac{1}{2})]\approx
h'(\frac{k}{N}-\frac{1}{2})(i-k)
\ee
Then the sum takes the form

\be\ba{l}
\log D\Big(\frac{k}{N}-\frac{1}{2}\Big)=
\Big({\dsp \sum_{|i-k|>M}+\sum_{|i-k|<M}}\Big)
\log\Big(1+\frac{1/N}{h(\frac{i}{N}-\frac{1}{2})-
h(\frac{k}{N}-\frac{1}{2})}\Big)\approx \vspace{1pc}
\\
\frac{1}{N}{\dsp \sum_{|i-k|>M}}\frac{1}{h(\frac{i}{N}-\frac{1}{2})-
h(\frac{k}{N}-\frac{1}{2})}+
{\dsp \sum_{|i-k|<M}}
\log\Big(1+\frac{1}{h'(\frac{k}{N}-\frac{1}{2})(i-k)}\Big)
\to \vspace{1pc}
\\
{\dsp \pvint} dy \frac{\rho(y)}{y-h}+
{\dsp \sum_{n=1}^{\infty}}\frac{[-\rho(h)]^n}{n}
{\dsp \sum_{q=1}^\infty}
\frac{1+(-1)^q}{q^n} = {\dsp \pvint} dy \frac{\rho(y)}{y-h}
+\log\Big(\frac{\sin \pi \rho(h)}{\pi \rho(h)}\Big)
\ea
\label{sum}
\ee

If $\frac{1}{2}<\rho(h)<1$, the analysis has to be more
careful. In this case, the product goes actually over columns of
the Young table and there are a lot of cancellations between the
numerator and the denominator in \eq{D}.
Let \mbox{$(j_1\equiv1,j_2,\ldots,j_m,j_{m+1}\equiv N+1)$}
be a sequence of
indices such that \mbox{$n_{j_k-1}-n_{j_k}=1$} and
$n_{j_k-2}-n_{j_k-1}=0$, then we have

\be
D\Big(\frac{j_k}{N}-\frac{1}{2}\Big)=(j_{k+1}-j_k)
\prod_{s\neq k}
\frac{j_{s+1}-j_k+s-k}{j_s-j_k+s-k}
\label{D2}
\ee
In this case, if $|s-k|<M$, we can expand
\mbox{$j_s-j_k\approx-\frac{\rho(h)}{1-\rho(h)}(s-k)$} and
analogously to \eq{sum} find

\[
\log D\Big(\frac{j_k}{N}-\frac{1}{2}\Big)\approx
{\dsp \frac{1}{N}\sum_{|i-k|>j_{k+M}-j_k}}
\frac{1}{h(\frac{i}{N}-\frac{1}{2})-h(\frac{j_k}{N}-\frac{1}{2})}+
\]\be
{\dsp \sum_{|s-k|<M}}\log\Big(1+\frac{\rho(h)}{s-k}\Big)
\to
{\dsp \pvint} dy \frac{\rho(y)}{y-h}
+\log\Big(\frac{\sin \pi \rho(h)}{\pi \rho(h)}\Big)
\label{sum2}
\ee
The factor $j_{k+1}-j_k\to \frac{\rho(h)}{1-\rho(h)}$ in \eq{D2}
combines with the density of vacancies (\ref{eta}) and we find the
solution which is valid for an arbitrary $\rho(h)$

\be
W(A_1,A_2) = \int \frac{dh}{\pi} \sin \pi \rho(h)
\exp\Big( - \pvint dy \frac{\rho(y)}{h-y} + A_2h\Big)
\label{W=int}
\ee
This formula can be nicely represented as the contour integral

\be
W(A_1,A_2) = \oint_C \frac{dh}{2\pi i} e^{A_2h-f(h)}
\label{W=}
\ee
where $f(h)$ is defined in \eq{f} and the contour $C$ encircles
the cut of $f(h)$. Obviously $W(A_1,0)=1$: to prove it one has to expand
the exponential in powers of $f(h)$ and take the residue at the infinity.
On the other hand, owing to the saddle-point equation (\ref{spe1}),
\eq{W=int} can be rewritten as

\be
W(A_1,A_2) = \int \frac{dh}{\pi} \sin \pi \rho(h)
e^{\frac{A_2-A_1}2 h}
= \oint_{\bar{C}} \frac{dh}{2\pi i} e^{-A_1h+f(h)}
\label{W=perm}
\ee
where the contour $\overline{C}$ encircles the cut of $f(h)$ in the
negative direction. Here, $W(0,A_2)=1$ is obvious.

\section{Particular cases}

In the weak coupling phase,
$f(h)=\frac{A}{2}h-\sqrt{\frac{A^2h^2}{4}-A}$ ($A=A_1+A_2$),
and it is convenient to
introduce the variable $z=\frac{\sqrt{A}}{if}$ in terms of which the integral
(\ref{W=}) takes the form

\be
W_{\rm wc}(A_1,A_2)=\frac{1}{i\sqrt{A}}\oint_C \frac{dz}{2\pi i}
\Big(1+\frac{1}{z^2}\Big) e^{iaz+i\frac{b}{z}}
\ee
where $a=\frac{1}{2}\sqrt{A}-\frac{A_1-A_2}{2\sqrt{A}}$ and
$b=\frac{1}{2}\sqrt{A}+\frac{A_1-A_2}{2\sqrt{A}}$. Expanding the
exponential and taking the residue at zero, we find

\be\ba{rl}
W_{\rm wc}(A_1,A_2)&=\frac{1}{i\sqrt{A}}{\dsp \oint_C} \frac{dz}{2\pi i}
\Big(1+\frac{1}{z^2}\Big) {\dsp \sum_{n=0}^{\infty} }
\frac{i^{2n+1}}{n!(n+1)!}(a^{n+1}b^nz+a^nb^{n+1}\frac{1}{z})=
\vspace{1pc}
\\
&{\dsp \sum_{n=0}^{\infty} }
\frac{(-1)^n}{n!(n+1)!}(ab)^n=\sqrt{\frac{A_1+A_2}{A_1A_2}}
J_1\Big(\sqrt{\frac{4A_1A_2}{A_1+A_2}}\Big)
\ea
\ee
So, we have reproduced the formula (\ref{W=J1}).

In the strong coupling phase, we have to use the Douglas-Kazakov solution
(\ref{DKsol}). It is convenient to change the variable

\be
h=b\, \sn u
\label{h=}
\ee
then we have

\be
f(b\,\sn u)=\frac{bA}{2}\sn u -2KZ(u)
\ee
where

\be
Z(u)= \frac{\d}{\d u} \log \theta_4\Big(\frac{u}{2K}\Big)=
\frac{4\pi}{2K}\sum_{m=1}^\infty \frac{q^m}{1-q^{2m}}
\sin\frac{m\pi u}{K}
\ee
is one of classical Jacobi's functions \cite{Erd}.
It is not double periodic

\be
Z(u+2K)=Z(u) \hspace{1cm} Z(u+i2K')=Z(u)-\frac{i\pi}{K}
\ee
However, we have the combination $\exp\{2KZ(u)\}$ which is elliptic.
Therefore, the contour integral can be reduced to a residue
at $u=iK'$ (which is mapped to the infinity by the function
(\ref{h=})).

\be
W_{\rm sc}(A_1,A_2)=b\oint_{C'} \frac{du}{2\pi i} \cn u\,\dn u\,
e^{\frac{A_2-A_1}{2}b\sn u + 2KZ(u)}
\label{intdu}
\ee

Let us notice that

\be
\frac{Ab}{2}\sn u -2KZ(u)=-4\pi\sum_{m=1}^\infty
(-1)^m\frac{q^{\frac{m}{2}}}{1-q^m}\sin\frac{m\pi u}{2K}
=-\frac{\theta'_3\Big(\frac{u}{4K}\Big|
\frac{\tau}{2}\Big)}{\theta_3\Big(\frac{u}{4K}\Big|\frac{\tau}{2}\Big)}
\ee
and

\be
\frac{Ab}{2}\sn u + 2KZ(u)=4\pi\sum_{m=1}^\infty
\frac{q^{\frac{m}{2}}}{1-q^m}
\sin\frac{m\pi u}{2K}
=\frac{\theta'_4\Big(\frac{u}{4K}\Big|
\frac{\tau}{2}\Big)}{\theta_4\Big(\frac{u}{4K}\Big|\frac{\tau}{2}\Big)}
\ee
where $\tau=i\frac{K'}{K}$ and $q=e^{i\pi\tau}$.
For convenience, let us shift the variable $u=z+iK'$, then

\be
\cn(z+iK')=\frac{\dn z}{i k\sn z}\hspace {1cm}
\dn(z+iK')=\frac{\cn z}{i \sn z}
\ee

\be
4K\frac{\d}{\d z}\log\theta_3\Big(\frac{z+iK'}{4K}\Big|\frac{\tau}{2}\Big)
=4K\frac{\d}{\d z}\log\theta_2\Big(\frac{z}{4K}\Big|\frac{\tau}{2}\Big)
-i\pi\stackrel{def}{=} \phi_2(z)-i\pi
\ee
and

\be
4K\frac{\d}{\d z}\log\theta_4\Big(\frac{z+iK'}{4K}\Big|\frac{\tau}{2}\Big)
=4K\frac{\d}{\d z}\log\theta_1\Big(\frac{z}{4K}\Big|\frac{\tau}{2}\Big)
-i\pi\stackrel{def}{=} \phi_1(z)-i\pi
\ee

Now we are able to write \eq{W=} in the strong coupling phase as the
integral

\be
W_{\rm sc}(A_1,A_2)=-a\oint_O \frac{dz}{2\pi i}
\frac{\cn z \, \dn z}{\sn^2 z} e^{\frac{A_1}{A}\phi_2(z) +
\frac{A_2}{A}\phi_1(z)}
\label{Wsc=}
\ee
where the contour $O$ encircles zero in the positive direction. Let us
give the Fourier expansions of $\phi_1$ and $\phi_2$:

\be
\phi_1(z)=\pi\cot \frac{\pi z}{4K} + 4\pi\sum_{m=1}^\infty
\frac{q^m}{1-q^m} \sin \frac{m\pi z}{2K}
\ee

\be
\phi_2(z)=-\pi\tan \frac{\pi z}{4K} + 4\pi\sum_{m=1}^\infty
(-1)^m\frac{q^m}{1-q^m} \sin \frac{m\pi z}{2K}
\ee

The Riemann surface of $f(h)$ has two sheets and, therefore, two points
in every mesh on the $z$-plane correspond to the infinity of
the $h$-plane ($4Kn+i2K'm$ and $4K(n+\frac12)+i2K'm$). We can as well
shrink the contour on the second sheet or, equivalently,
around the point $2K$. As
$\phi_1(z+2K)=\phi_2(z)$ and $\phi_2(z+2K)=\phi_1(z)$, we find that
$W(A_1,A_2)=W(A_2,A_1)$ holds in the strong coupling phase as well.

The representation (\ref{Wsc=}) is convenient for the investigation of
the transition point, $A=\pi^2$, which corresponds to the limit
$k\to 0$, $K'\to \infty$, $K\to\frac\pi2$ when

\be
\sn z \to \sin z \hspace{1cm} \cn z \to \cos z \hspace{1cm} \dn z \to 1
\ee
and we find

\[
W_{\rm sc}(A_1,\pi^2-A_1)=-\frac{2}{\pi}\oint_O \frac{dz}{2\pi i}
\frac{\cos z}{\sin^2 z} e^{-\frac{A_1}\pi\tan\frac{z}2+(\pi-\frac{A_1}{\pi})
\cot\frac{z}2} = \]
\[
\frac{1}{\pi}\oint_O \frac{dx}{2\pi i}
\Big(1-\frac1{x^2}\Big)e^{-\frac{A_1}\pi x + (1-\frac{A_1}{\pi})\frac1x}
=\sum_{n=0}^\infty\frac{(-1)^n}{n!(n+1)!}
\Big(\frac{A_1(\pi^2-A_1)}{\pi^2}\Big)^n \]\be
=\frac{\pi^2}{A_1(\pi^2-A_1)}
J_1(2\sqrt{A_1(1-A_1/\pi^2)})
\ee
which coincides with the weak coupling solution at the transition point.

The Wilson loop on the plane ($A_1\to\infty$, $A_2$ finite)
corresponds to the limit $k\to1$, $K'\to\frac\pi2$, $K\to\infty$. Here,
it is convenient to make the modular transformation $\tilde{k}=k'$,
$\tilde{k'}=k$, $\widetilde{K}=K'$, $\widetilde{K'}=K$, $z=-iu$ after which

\be
\sn(z,k)=\frac{\sn(u,k')}{i\cn(u,k')}\hspace{1cm}
\cn(z,k)=\frac{1}{\cn(u,k')}\hspace{1cm}
\dn(z,k)=\frac{\dn(u,k')}{\cn(u,k')}
\ee
and
\be\ba{l}
\phi_1(z)=4Ki\frac{\d}{\d u}\log\theta_1(\frac{u}{2K'}\frac1{2\tilde{\tau}}|
-\frac1{2\tilde{\tau}})=
4Ki\frac{\d}{\d u}\log\theta_1(\frac{u}{2K'}|2\tilde{\tau}) +
i\frac{\pi u}{K'}\vspace{1pc}
\\
\phi_2(z)=4Ki\frac{\d}{\d u}\log\theta_2(\frac{u}{2K'}\frac1{2\tilde{\tau}}|
-\frac1{2\tilde{\tau}})=
4Ki\frac{\d}{\d u}\log\theta_4(\frac{u}{2K'}|2\tilde{\tau}) +
i\frac{\pi u}{K'}
\ea
\ee
where $\tilde{\tau}=i\frac{K}{K'}$.
After the modular transformation the integral takes the form

\be
W_{\rm sc}(A_1,A_2)=-ai\oint_O \frac{du}{2\pi i}
\frac{\dn(u,k')}{\sn^2(u,k')}
e^{\frac{2iKA_1}{K'A}\tilde{\phi}_4(u) +
\frac{2iKA_2}{K'A}\tilde{\phi}_1(u)+i\frac{\pi u}{K'}}
\ee
where $\tilde{\phi}_n(u)=
2K'\frac{\d}{\d u}\log\theta_n(\frac{u}{2K'}|2\tilde{\tau})$.

The natural expansion parameter here is

\be
\qt\equiv e^{i\pi \tilde{\tau}}=\exp\Big( -\frac{\pi a A}{4K'}\Big)
\to e^{-\frac{A}4}
\ee
and, to perform the expansion, it is convenient to reexpress all
quantities involved through the theta functions:

\be\ba{l}
\dsp
\frac{2K'}{\pi}=\theta_3^2(0) = 1+4\qt+4\qt^2+\ldots \nl
\frac{2K'}{\pi}\frac{\dn(2K'v,k')}{\sn^2(2K'v,k')}=
\frac{\theta_2^2(0)\theta_4(0)}{\theta_3(0)}
\frac{\theta_3(v)\theta_4(v)}{\theta_1^2(v)}=
\frac1{\sin^2\pi v}[1-4\qt-4\qt^2(\cos 2\pi v - \sin^2 2\pi v)
+\ldots]\nl
a=\frac1{2E-k'^2K}=\frac1{2(1-4\qt+12\qt^2-8\qt^2\log\qt+\ldots)} \nl
\tilde{\phi}_1(2K'v)=\pi\cot\pi v + 4\pi \qt^4\sin 2\pi v+\ldots \nl
\tilde{\phi}_4(2K'v)= 4\pi \qt^2\sin 2\pi v+\ldots
\ea\ee
We have also to
use the relation $\frac{2K}{AK'}\to \frac1{2\pi}$,
which follows from \eq{eqspar} in the $K\to\infty$ limit.
Then, we find

\be\ba{l}\dsp
W_{\rm sc}(A_1,A_2)=-i\pi a\oint \frac{dv}{2\pi i}
e^{i\frac{\pi aA_2}{2K'}\cot\pi v + 2\pi i v} \nl
\Big\{1-4\qt+i\frac{2\pi aA_1}{K'}\qt^2\sin 2\pi v
-4\qt^2(\cos2\pi v - \sin^2 2\pi v)+\ldots\Big\}= \nl
{\dsp
-\frac12 \oint \frac{dx}{2\pi i}\frac{x-1}{x+1}}
e^{\frac{\pi aA_2}{2K'}x}
\Big\{1+4\qt^2+8\qt^2\log\qt -2A_1\qt^2\frac{x}{x^2-1}- \nl
4\qt^2\Big[\frac{x^2+1}{x^2-1}+
\Big(\frac{2x}{x^2-1}\Big)^2\Big]
+\ldots\Big\}
\ea\ee
The integral is reduced to residues at $x=1$ and $x=-1$ and we find

\be
W_{\rm sc}(A_1,A_2)=
e^{-\frac{A_2}2} + e^{-\frac{A_1}2} +(-1+3A_2+\frac{A_2^2}2)
e^{-\frac{2A_2+A_1}2} + \ldots
\label{Wsc=exp}
\ee

As the points $0$ and $2K$ are separated in the $K\to\infty$ limit by
the infinite distance, we have the non-symmetric expansion. If we
shifted the variable \mbox{$z\to z+2K$} before making the modular
transformation, we would find the answer with permuted $A_1$ and $A_2$.

\bigskip

The first look at the expansion (\ref{Wsc=exp}) shows that it is
compatible with the sum-over-branched-coverings interpretation.
The first terms show the standard $QCD_2$ area low. The last one
represents the simplest non-trivial covering with a disc.
Nevertheless, there is some
puzzle here which reads: ``How can the property $W(0,A)=W(A,0)=1$ be agreed
with an open-string picture, we would like to have?''.

\bigskip
{\large\bf Acknowledgments}

\medskip
The discussions with V.A.Kazakov have very much contributed to this work.
I would like to thank B.Rusakov for the numerous stimulating discussions
and kind hospitality at Physics Department of the Tel-Aviv University,
where this work was started.

\bigskip

{\bf\large Note added}: After having completed the manuscript, I learned
that \eq{W=} has also been obtained by Jean-Mark Daul and Volodya
Kazakov \cite{DaulKaz}.

\end{document}